\documentclass[superscriptaddress,twocolumn,floatfix,aps,preprintnumbers]{revtex4}

\usepackage{graphicx,color}
\usepackage{amsmath,amssymb}
\usepackage{natbib}
\usepackage{color}
\usepackage[normalem]{ulem}
\usepackage{silence}
\usepackage[colorlinks=true,linkcolor=blue,citecolor=blue,urlcolor=blue,hypertexnames=false]{hyperref}
\usepackage[caption=false]{subfig}
\usepackage{graphics,amssymb,amsmath,epsfig,color}

\newcommand{\KL}{\text{KL}}

\begin{document}

\title{Beyond Explained Variance: A Cautionary Tale of PCA}

\author{Gionni Marchetti}
\affiliation{Barcelona, Spain}

\email{gionnimarchetti@gmail.com}

\date{\today}

\begin{abstract}
We address shortcomings of principal component analysis (PCA) for visualizing high-dimensional data lying on a nonlinear low-dimensional manifold via two-dimensional scatterplots, focusing on a fossil teeth dataset from the early mammalian insectivore Kuehneotherium ($n_s = 88$, $n = 9$). While the PCA scatter plot reported by Jolliffe and Cadima (Philosophical Transactions of the Royal Society A, 2016) shows clustering in the region where $\mathrm{PC2} < 0$, our analysis based on $t$-SNE and persistent homology (PH) reveals a ring-like structure with no evident clustering and intrinsic dimensionality equal to one. We further propose a generative probabilistic–geometric model in which the data are sampled uniformly from a unit circle. Under this model, pairwise cosine distances follow an arcsine distribution, in qualitative agreement with the observed U-shaped distribution, thereby independently supporting the analysis based on $t$-SNE and persistent homology.
\end{abstract}

\maketitle

\section{Introduction}\label{Intro}

Principal component analysis (PCA) is a widely used tool in data analysis and machine learning (ML), employed for feature extraction, dimensionality reduction, and data visualization~\cite{Pearson1901, Hotelling1933, jolliffe2002pca, shlens2014, Greenacre2022}. Owing to its simple mathematical formulation, relatively low computational cost, and the increasing availability of data, including synthetic datasets in materials science~\cite{scheidgen2023nomad, Horton2025} and large-scale biological data~\cite{berman2000protein, varadi2024alphafold}, its use is expected to grow substantially in the coming years.

In particular, PCA is routinely used for visualizing high-dimensional data as points in a two-dimensional scatterplot, where it can reveal clustering structures and potential outliers in the original high-dimensional dataset. However, as a linear method, PCA can produce misleading visualizations when the data under study has a nonlinear structure. In fact, PCA implicitly assumes that the data lie near an optimal hyperplane of rank $m$ (or, equivalently, a linear subspace of the same dimension, assuming the data has been standardized)~\cite{hastie2017, Deisenroth2020}, while the data actually may lie on a nonlinear manifold. 

Therefore, it would be overly simplistic to assume that strong PCA performance in terms of explained variance—e.g., when the first two principal components ($\mathrm{PC1}$ and $\mathrm{PC2}$) account for large variance, necessarily implies that the corresponding two-dimensional scatterplot provides a meaningful representation of the underlying structure of the high-dimensional data.

That being said, this work aims to address this potential shortcoming by analyzing a dataset of fossil teeth from the early mammalian insectivore \emph{Kuehneotherium}, one of the earliest known mammals, which lived approximately $200$ million years ago, during the Late Triassic to Early Jurassic epochs. This data consists of $n_s = 88$ samples from Mesozoic fissure fillings in South Wales, each described by $n = 9$ linear measurements (or equivalently variables)~\cite{Gill2014} (see Sect.~\ref{dataset} for details). Therefore, these data points belong to $\mathbb{R}^{9}$.  We chose this specific data because Ref.~\cite{Joliffe2016} presented a two-dimensional scatterplot (see panel a of Fig.~\ref{fig:fig1} and also Fig.~\ref{fig:fig2_SM}), for which the data tend to cluster in the region where $\mathrm{PC2} <0$ is negative, as a paradigmatic example of data visualization using PCA, with the first two principal components accounting for approximately $95\%$ of the total variance. 

In contrast, a reassessment of the PCA pattern shows that the two-dimensional embedding obtained using $t$-distributed stochastic neighbor embedding ($t$-SNE)~\cite{vandermaaten2008, Kobak2019}, shown in panel b of Fig.~\ref{fig:fig1} (see also Fig.~\ref{fig:fig4_SM}), suggests that the high-dimensional data exhibit a ring-like structure. To further support the existence of this topological feature, we compute the corresponding persistent homology (PH) diagrams, a key tool in topological data analysis (TDA)~\cite{Carlsson2009, Otter2017, Wasserman2018, Munch_2017, chazal2021}. These persistence diagrams, obtained using the Vietoris–Rips simplicial complex based on Euclidean and cosine distances, are shown in panels c and d, respectively,  in Fig.~\ref{fig:fig1}. Both indicate the presence of a loop; however, only the diagram based on the cosine distance provides strong evidence for its persistence.

Accordingly, we show that the choice of distance (or metric) is critical for revealing the underlying topological and geometric features of the data, as previously noted in Ref.~\cite{chazal2021}. In particular, the Euclidean distance can hinder the proper functioning of $t$-SNE and persistent homology due to its sensitivity to measurement noise and the relatively high dimensionality of the data~\cite{Vershynin2018, damrich2024}. In contrast, the cosine distance provides a more suitable similarity measure in this case, as it captures the circular symmetry of the data.

Finally, we propose a probabilistic–geometric modeling approach after estimating the intrinsic dimensionality (ID) of the data using various methods, which yields a value of one. Accordingly, we model the data as points lying on a unit circle, sampled uniformly at random. Under this assumption, their pairwise cosine distances follow a U-shaped arcsine probability density function (pdf) (see Eqs.~\ref{eq:f_D},~\ref{eq:arcsine})~\cite{Levy1939}. Such a model, based on the circular symmetry and intrinsic dimensionality of the data, predicts that the distribution of cosine distances should exhibit a U-shaped histogram. This bimodal pattern is indeed observed in the empirical distribution, although the experimental data points neither lie exactly on a circle nor are sampled uniformly at random. This finding further supports the presence of a loop-like structure in the high-dimensional data.

\section{Methods}\label{Methods}

To begin, we refer the reader to Section~\ref{dataset} of Supplemental Material (SM), for details on the fossil tooth dataset $\mathcal{X} = \{\mathbf{x}_1, \mathbf{x}_2, \cdots, \mathbf{x}_i, \cdots, \mathbf{x}_{n_s}\}$  and its standardization. 
It suffices to note that the standardized data are stored in an $n_s \times n$ data matrix, denoted by $X^{\ast}$.

\subsection{Principal Component Analysis}\label{PCA}

Standard principal component analysis can be understood in two equivalent ways. Accordingly, finding a low-dimensional representation of a given high-dimensional dataset $\mathcal{X}$, can be formulated either as preserving as much of the total variance as possible, or as orthogonally projecting each data point $\mathbf{x}_i$ onto a new point $\mathbf{\tilde{x}}_i$ that lies in an optimal subspace, thereby minimizing the (average) reconstruction error~\cite{Deisenroth2020}. However, only the latter formulation, which explicitly considers an $m$-rank linear modeling for representing the data, highlights that PCA is indeed a linear algorithm~\cite{hastie2017}.

Overall, PCA boils down to computing the eigenvalues  $\lambda_i$ ($\lambda_1 \geq \lambda_2 \geq \cdots \lambda_i \geq \cdots \lambda_n$), each corresponding to the explained (or preserved) variance along the $i$-th principal component axis, of the sample covariance matrix $S=\left( n_s -1 \right)^{-1} X^{\ast^{T}} X^{\ast}$. This task is performed efficiently by using the singular value decomposition (SVD)~\cite{strang1993, stewart1993, Shinn2023}. Using SVD, one finds  $X^{\ast}=WLV^T$ where $W$ and $V$ are two orthogonal matrices, and $L$ is a diagonal matrix~\cite{Joliffe2016, Greenacre2022}. Consequently, the singular values $s_i$  ($s_{1}^{2} \geq s_{2}^{2}  \geq \cdots \geq s_{n}^{2} \geq 0 $) correspond the diagonal entries of $L$, each satisfying $\lambda_i = \left(n_s -1 \right)^{-1} s_{i}^{2}$. Next, assuming that $m$ is the dimension of the linear subspace on which the data points are projected, the corresponding reconstruction error is $J_m = n_s^{-1}\sum_{j= 1}^{n_s} \lVert \mathbf{x_j} - \mathbf{\tilde{x}_j} \rVert_2^{2} $ which is equivalent to $J_m = \sum_{l= m + 1}^{n} \lambda_l$~\cite{Deisenroth2020}. Note that in this work, PCA is computed using the scikit-learn library~\cite{pedregosa2011}.

\subsubsection{Estimating Intrinsic Dimensionality using PCA}

Within PCA, the methods for estimating the intrinsic (or effective) dimension, which, in this context, corresponds to deciding how many principal components (PCs) to keep. To this end, various heuristics exist. 

In this work, we shall consider the following ones: Kaiser criterion (also known as the Kaiser-Gutman rule)~\cite{Kaiser1960, jolliffe2002pca}, the elbow method~\cite{tenenbaum2000, Geron2017}, the participation ratio (PR)~\cite{Kramer1993, RECANATESI2022}, and finally Gavish-Donoho (GD) optimal hard threshold~\cite{Gavish2014, Brunton2019}.

According to the Kaiser criterion,  the number of PCs that capture most of the variance corresponds to those for which $\lambda_i \geq 0.7$. The elbow method relies on determining the elbow (or knee) of the reconstruction error curves, that is, $J_m$ as a function of the dimension $m$ of the optimal subspace. The PR estimates the number of dimensions along which the data spreads according to the formula $D_{\rm PR} = \left( \sum\limits_{i=1}^{n} \lambda_i \right)^2/\sum\limits_{i=1}^{n} \lambda_i^2$. Finally, in the DH approach,  the ID is estimated by discarding singular values equal to or greater than a threshold $\tau^{\ast} $, i.e., $s_i \geq \tau^{\ast}$. 
We refer the reader to Sec.~\ref{Gavish-Donoho}, where it is explained how the computed $\tau^{\ast}$, which depends on the Mar\v{c}enko--Pastur (MP) distribution~\cite{marcenko1967}, is derived assuming that the noise is unknown.

\subsection{t-Distributed Stochastic  Neighbor Embedding }\label{tSNE}

In contrast to PCA, $t$-SNE aims to construct a low-dimensional embedding (or map) $\mathcal{Y} = \{\mathbf{y}_1, \mathbf{y}_2, \cdots, \mathbf{y}_{\rm n_s} \}$ such that high-dimensional neighbors remain neighbors in the embedding. This method renounces the preservation of the pairwise distances, thereby avoiding the possible issues, e.g., the norm concentration,  arising from the high-dimensionality of the data (the curse of dimensionality)~\cite{debodt2025}
Denoting with $P$ and $Q$ the joint-probability distributions associated to $\mathcal{X}$ and $\mathcal{Y}$, respectively, $t$-SNE looks for 
the low-dimensional embedding, by   minimizing, through the gradient descent, the  Kullback-Leibler (KL) divergence $\KL(P\|Q)$ between $P$ and $Q$, whose expression is given by
\begin{equation}\label{eq:KL}
	\KL(P\|Q)=\sum_{i=1}^{n_s}  \sum_{j=1, j \neq i }^{n_s}  p_{i j} \log \frac{p_{i j}}{q_{i j}} \, ,
\end{equation}
where the symmetric probabilities $p_{ij}$ and  $p_{ij}$ are defined in Sec.~\ref{tSNE_details}. 

This manifold reduction algorithm depends on a hyperparameter: the perplexity parameter $\tau_p$.
The latter can vary from $5$ to $50$, $30$ being the default value~\cite{vandermaaten2008, Kobak2019}. Accordingly, there is no principled way to find its optimal value. Furthermore, it is usually applied to large datasets, which is not the present case. Note that in this work,  $t$-SNE is computed using \textsf{openTSNE}~\cite{policar2024}.

\subsection{Persistent Homology}\label{homology}

Persistent homology diagrams are used to detect topological features of data across a scale parameter, which in our case is the diameter of (filtered) Vietoris–Rips simplicial complex. As this parameter increases, the data progressively thickens, and topological features appear (birth) and disappear (death), giving rise to points in the persistent diagram. The difference between death and birth defines the persistence (or lifetime) of a feature, with longer lifetimes indicating more significant structures. In contrast, short-lived features located near the diagonal of the persistent diagram are attributed to noise. 

The similarity between these diagrams can be conveniently measured by various metrics, e.g., the bottleneck distance~\cite{Munch_2017}  and the Wasserstein distances~\cite{Villani2008, PeyreCuturi2018}.

The typical topological features considered are $H_0$, $H_1$, and $H_2$, corresponding to the number of connected components, loops (or holes), and voids, respectively. In what follows, persistent homology diagrams, along with their equivalent persistent barcodes, and the bottleneck distance are computed using the giotto-tda toolbox~\cite{tauzin2020giottotda}.

\subsection{Cosine Distance}\label{distances}

This works crucially depends on the choice of the appropriate distance. In fact, it is found that replacing the Euclidean distance  by the cosine distance $d_{\rm cos}$,  $t$-SNE and persistent homology yield reasonable consistent results. That being said, the cosine distance reads~\cite{Murphy2012}
\begin{equation}\label{eq:cosine_distance}
d_{\rm cos}\left(\mathbf{x}_i ,  \mathbf{x}_j \right) = 1 - \frac{ \mathbf{x}_i  \cdotp \mathbf{x}_j}{\lVert  \mathbf{x}_i    \rVert_2 \lVert  \mathbf{x}_j \rVert_2 }  \, .
\end{equation}

Note that Eq.~\ref{eq:cosine_distance} can also be written as $d_{\rm cos} = 1 - \cos\theta$  where $\theta \in [0, \pi]$ is the angle between $\mathbf{x}_i$ and $\mathbf{x}_j$. Therefore, this metric does not depend on the magnitudes of the data points, but only on the angle between them.

\section{Results and Discussion}\label{results}

\begin{figure*}[t]
\centering
\includegraphics[width=\textwidth]{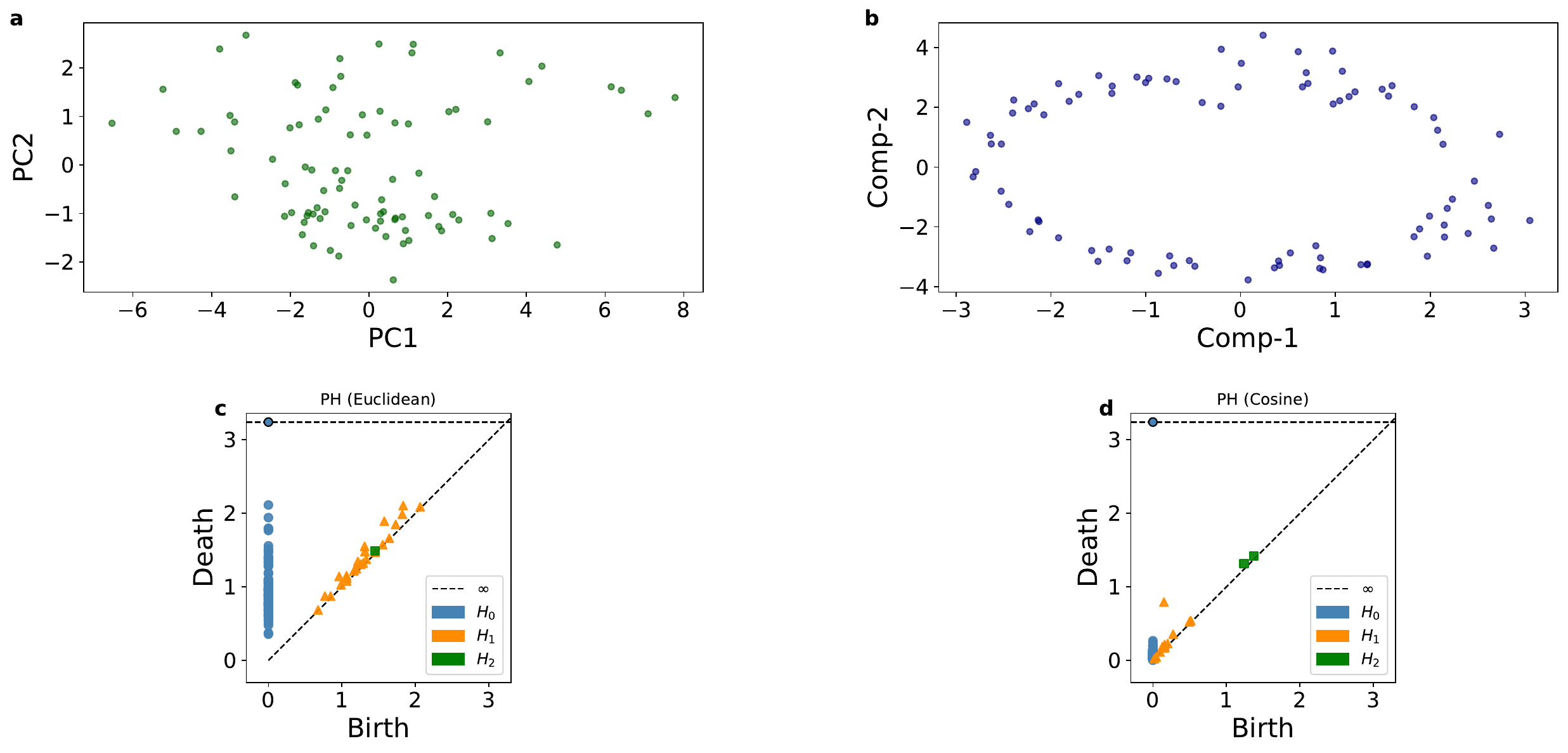}
\caption{All computations are performed on the standardized fossil teeth data. \textbf{a})~PCA two-dimensional optimal subspace. 
\textbf{b})~$t$-SNE two-dimensional embedding obtained using cosine distance
and PCA initialization ($\tau_p = 29$). 
\textbf{c})~Persistent homology diagram computed using Vietoris-Rips complex and Euclidean distance. 
\textbf{d})~Persistent homology diagram computed using Vietoris-Rips complex and cosine distance.}
\label{fig:fig1}
\end{figure*}

In panel a) of Figure~\ref{fig:fig1}, we show the two-dimensional scatterplot of the fossil data obtained using PCA. This pattern is in excellent agreement with that reported by Jolliffe and Cadima, obtained using the statistical software R~\cite{Joliffe2016}. 
Regarding PCA performance in terms of explained variance, we find that PC1 and PC2 account for $74.9\%$ and $18.8\%$ of the total variance, respectively. By comparison, the corresponding values reported using R~\cite{Joliffe2016} are $78.8\%$ and $16.7\%$. Despite an overall difference of approximately $1.8\%$, both analyses yield strong performance, with the first two principal components explaining approximately $93.7\%$–$95.5\%$ of the total variance.

We note a tendency toward clustering in the region of the principal subspace where PC2 is negative. However, this contrasts with what Andrews plots predict since the curves do not form bands (see Fig.~\ref{fig:fig1_SM} and Sec.~\ref{Andrews} for details). As a result, we do not expect to find clustering in the high-dimensional data.

In panel b of Figure~\ref{fig:fig1}, a two-dimensional embedding obtained using $t$-SNE is shown. It is computed with PCA initialization to better preserve the global structure of the data and uses the cosine distance, which is less affected by the curse of dimensionality. There is no a priori rule for selecting the perplexity parameter, and $t$-SNE is typically applied to much larger datasets. Therefore, the embedding shown in Figure~\ref{fig:fig1} corresponds to a somewhat subjective choice of $\tau_p = 29$, noting that ring-like structures emerge for all $\tau_p \geq 12$. Nevertheless, the presence of such a loop in the high-dimensional data is supported by the subsequent analysis. It is also worth noting that if the Euclidean distance were used instead in $t$-SNE, such a structure would not be detected (see, for instance, Fig.~\ref{fig:fig2_SM}).

Furthermore, the emergence of a ring-like structure, in contrast to the clustering tendency observed in PCA, does not contradict the Andrews plots.

Next, to further investigate the existence of a loop in the high-dimensional data, we resort to the persistent diagrams. In panels c) and d) of Fig.~\ref{fig:fig1}, the diagrams computed using Euclidean and cosine distances are shown. In both cases, the $H_2$ features (green squares) lie close to the diagonal, indicating low persistence and suggesting the absence of meaningful voids, with these features likely attributable to noise~\cite{Otter2017}. The behavior of the $H_0$ features is consistent across both metrics, with a single dominant long-lived component indicating that the data ultimately forms one connected component.

In the context of class $H_0$, we further observe that the features computed using cosine distance exhibit shorter lifetimes, reflecting the metric’s emphasis on angular similarity,  leading to a reduction in the range of filtration scales on which the features persist. This is further illustrated by the corresponding barcode representations; see Figs.~\ref{fig:fig6_SM} and~\ref{fig:fig7_SM}, together with the related analysis in Sec.~\ref{extra_results}.

For $H_1$, the cosine distance diagram shows a feature with relatively high persistence (the point farthest from the diagonal), suggesting the presence of a loop. In contrast, the Euclidean diagram exhibits several persistent features of moderate lifetime. Their difference is clarified by comparing their persistent barcodes, shown in  Figs.~\ref{fig:fig5_SM} and~\ref{fig:fig6_SM}. Note that 
using the cosine distance, a single dominant $H_1$ class emerges, while the remaining features are short-lived. This provides evidence for a loop structure, although its persistence depends on the choice of metric.

To quantify these differences, we compute the bottleneck distances between the persistence diagrams for each equivalence class, using the giotto-tda toolbox. We obtain $d_B^{(0)} = 1.0563$, $d_B^{(1)} = 0.3211$, and $d_B^{(2)} = 0.0381$ for $H_0$, $H_1$ and $H_2$, respectively. The relatively large value for $H_0$ reflects substantial differences in the merging scales of connected components, consistent with the shorter lifetimes observed under cosine distance. The small value for $H_2$ confirms that both metrics agree on the absence of voids. The intermediate value for $H_1$ indicates that while both metrics capture loop-like structures, their persistence differs, with the cosine distance more clearly isolating a single dominant feature. 

Next, another important feature of the data is its intrinsic dimensionality. In the present case, the ID can be inferred by exploiting the fact that PCA tends to overestimate it when the data lie on a nonlinear manifold. Specifically, if the data lie on an $m$-dimensional nonlinear manifold, PCA will typically estimate the dimensionality to be at least $m+1$~\cite{Zeng_2024}. A classic example is the synthetic Swiss roll dataset, for which PCA, using the elbow method, estimates the dimensionality as $m^{\ast}=3$, whereas the data actually lie on a two-dimensional manifold, i.e., $m^{\ast}=2$~\cite{tenenbaum2000}.

In this regard, the elbow method applied to the reconstruction error curve $J_m$ yields $m^{\ast} = 2$ for the fossil teeth data (see Fig.~\ref{fig:fig5_SM}, panel a). Similarly, only the first two singular values ($s_1\approx 24.36$, $s_2\approx 12.2$, $s_3\approx 4.63$), exceed the Gavish–Donoho optimal hard threshold $\tau^{\ast} \approx 4.75$, as illustrated in Fig.~\ref{fig:fig5_SM} (panel c), again indicating $m^{\ast} = 2$. The Kaiser and PR criteria are consistent with these results, also yielding $m^{\ast} = 2$ ($\lambda_1 \approx 6.82$, $\lambda_2 \approx 1.71$, and $\sum_{i=1}^{n} \lambda_i \approx 9.1$). However, given the nonlinear structure of the data, this value should be interpreted as an upper bound, leading to the conclusion that the data lie on a one-dimensional curved manifold, i.e., $m^{\ast} = 1$.

In the following, we combine this estimate of the intrinsic dimensionality with the observed symmetry of the ring-like structure to model the data from a probabilistic–geometric perspective. The aim is to show that such a generative model can qualitatively reproduce a characteristic pattern of the data, thereby further supporting our analysis independently of the data-driven algorithms $t$-SNE and persistent homology.

\begin{figure}
\centering
\includegraphics[width=8cm]{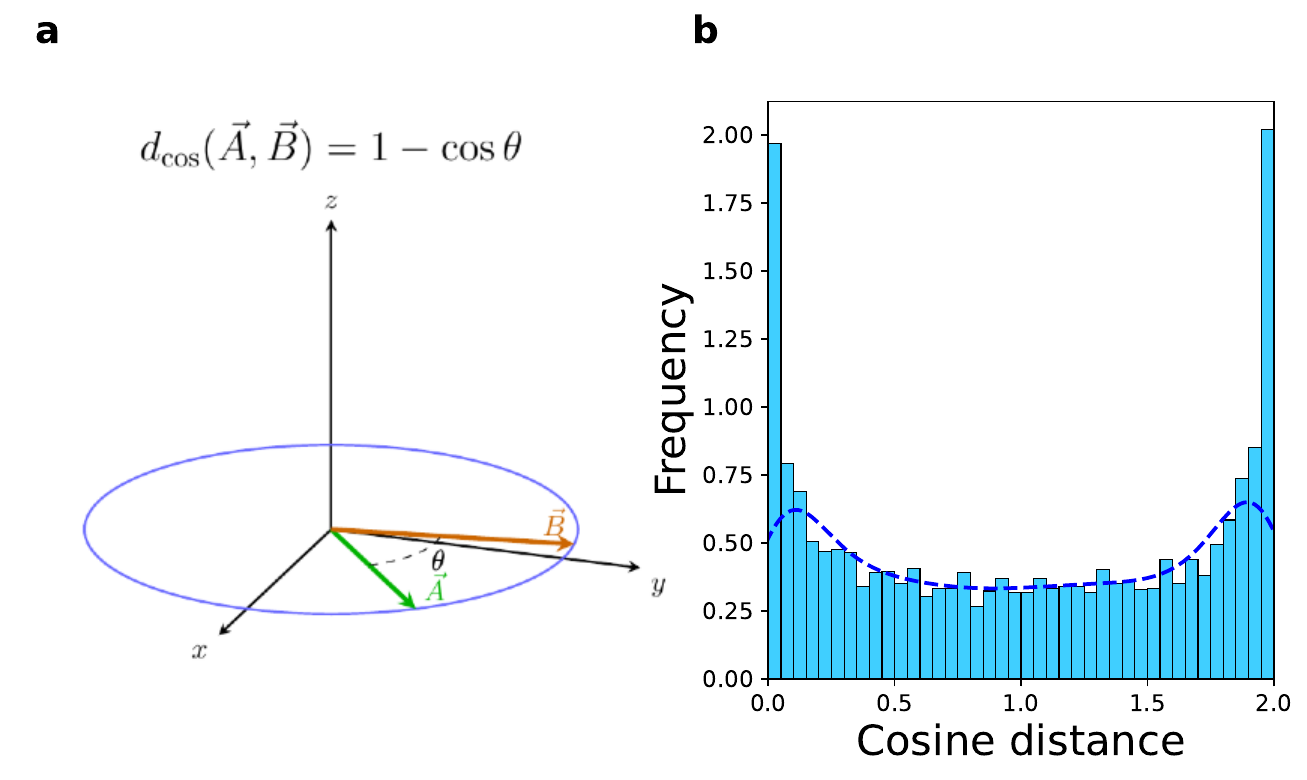}
\caption{\textbf{a})~Unit circle in $\mathbb{R}^{3}$, representing the geometric setup used for our generative probabilistic-geometric model.
\textbf{b})~ Histogram of $3828$ pairwise cosine distances of points on a unit circle in $\mathbb{R}^{9}$, generated uniformly at random.  The curve (dashed line) corresponds to the kernel density estimation (KDE).}
\label{fig:fig2}
\end{figure}

Accordingly, we model the fossil teeth data as a collection of data points lying on a unit circle embedded in a high-dimensional space. This idealized model encodes the previously identified intrinsic dimensionality and symmetry of the data, and is illustrated for the three-dimensional case in panel a of Fig.~\ref{fig:fig2}.

We then regard the fossil teeth data as points sampled uniformly at random from a circle. Consequently, the probability density function of the angular variable is given by $\theta \sim \mathrm{Uniform}(0,2\pi)$~\cite{MardiaJupp2000}:

\begin{equation}~\label{eq: uniform_pdf}
f_{\Theta}(\theta) = \frac{1}{2\pi}  \, ,
\end{equation}
with $\theta \in [0, 2\pi)$. Now, let us consider the cosine distance $d_{\rm cos}$ as a random variable $D$ defined according to $D = g\left(\Theta\right)$ where the function $y= g\left(\theta\right) = 1 - \cos \theta$ ($y \in [1, 2]$). Accordingly, the probability density function $f_{D}$ of the random variable $D$ can be derived using the change-of-variable formula for the non-monotonic transformation $g$. In such a case, this formula is~\cite{Casella2002}

\begin{equation}~\label{eq:change_of_variable}
f_{D}(y) = \sum_{i=1}^{2} f_{\Theta} \left(g_{i}^{-1}\left(y\right)\right)\left|\frac{d}{dy} g_{i}^{-1}\left(y\right) \right|\, .
\end{equation}
where the monotonic functions $g_1$ and $g_2$ correspond to the function $g$ restricted to intervals $[0, \pi)$ and  $[\pi, 2\pi)$, respectively. Their inverse functions are  $g_1^{-1}(y) = \arccos(1 - y)$ and $g_2^{-1}(y) = 2\pi - \arccos(1 - y)$. Due to the absolute value in Eq.~\ref{eq:change_of_variable}, each derivative of the inverse functions  $g_1^{-1}$ and $g_2^{-1}$ outputs $1/\sqrt{1-(1-y)^2}$. As a result, from Eq.~\ref{eq:change_of_variable} one obtains the following probability density function of random variable $D$: \begin{equation}\label{eq:f_D}
    f_D(y) = \frac{1}{\pi\sqrt{y (2- y)}}  \, ,
\end{equation}  
 with $y \in (0, 2)$. Next, the U-shaped arcsine distribution $f\left(x\right)$ with $x\in \left(0,1\right)$ has the following pdf~\cite{Levy1939, Bury1999}
\begin{equation}\label{eq:arcsine}
    f(x) = \frac{1}{\pi\sqrt{x(1-x)}}  \, ,
\end{equation}
 with $y \in (0, 1)$. By comparison of Eq.~\ref{eq:f_D} with  Eq.~\ref{eq:arcsine}, it is found that $f_D$ is nothing but the arcsine distribution stretched to the interval $(0,2)$. Such a distribution is U-shaped. This fact is illustrated in panel b of Fig.~\ref{fig:fig2} in which the histogram of cosine distances is generated by a numerical simulation of our proposed model using $88$ data points on the unit circle in $\mathbb{R}^{9}$, uniformly sampled at random. Note that data tends to pile up substantially on the interval ends, aligning with the expected divergence of arcsine distribution in those points (Eq.~\ref{eq:f_D}). 
The U-shape of this histogram is additionally shown by the curve corresponding to the kernel density estimation (KDE) computed using the SciPy library~\cite{mckinney-proc-scipy-2010}.

\begin{figure}
\centering
\includegraphics[width=8cm]{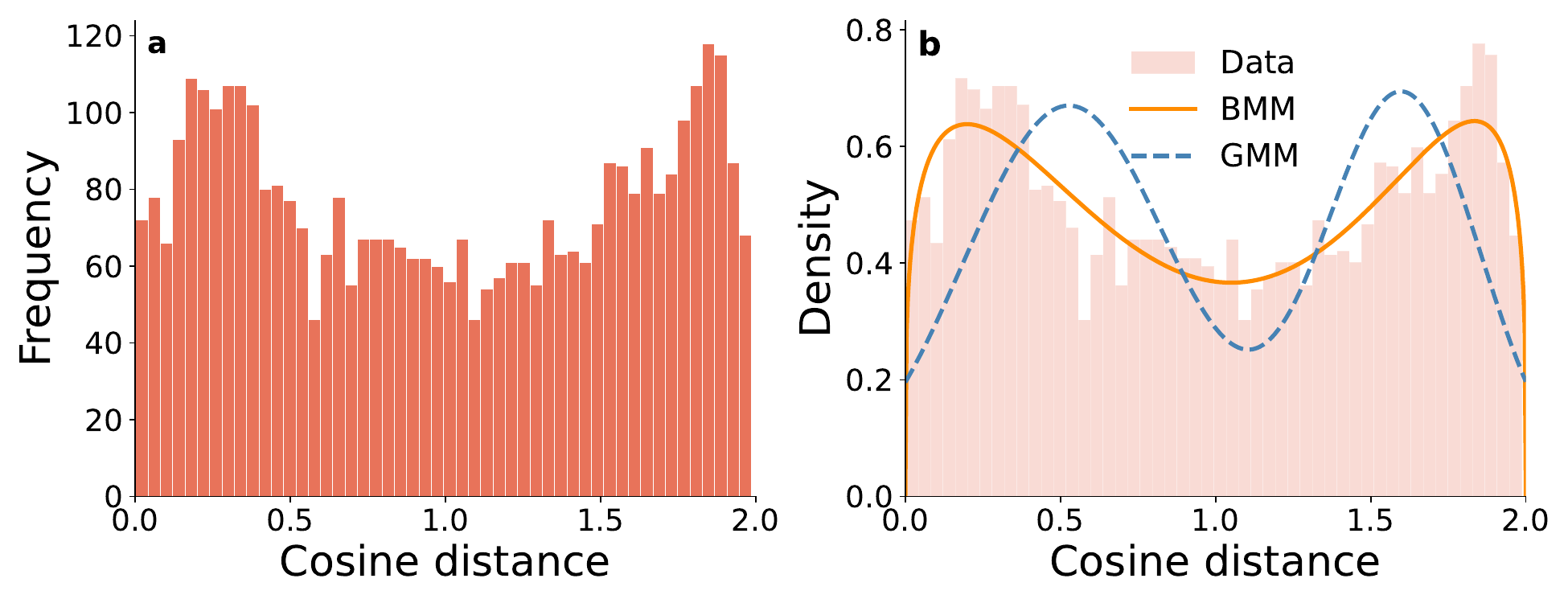}
\caption{\textbf{a})~Empirical histogram of the  $3828$ pairwise cosine distances between the data points belonging to fossil teeth data. 
\textbf{b})~ Fits of the cosine distance data using Beta and Gaussian  Mixture Models.}
\label{fig:fig3}
\end{figure}

Next, we compute the empirical histogram of the pairwise cosine distances from the samples belonging to the fossil teeth dataset (see panel a of Fig.~\ref{fig:fig3}). The distribution appears to be bimodal, as expected, quite similar to the KDE of the synthetic data obtained from the generative model. To make the above observation  more evident, we resort to fitting the empirical data using the two-component Gaussian mixture model (GMM) and two-component Beta mixture model (BMM), the Beta  being the following pdf $f(x; a, b)$ ($a, b > 0$)~\cite{Bury1999}
\begin{equation}
f(x; a, b) = \frac{1}{B(a,b)} \, x^{a-1} (1 - x)^{b-1}  \, ,
\end{equation}
where $B(a,b) = \Gamma(a)\,\Gamma(b)/\Gamma(a+b)$, $\Gamma\left(z\right)$ being the Gamma function and $x \in (0,1)$. We note in passing that the arcsine distribution is a special case of the Beta distribution, i.e., when $a=b=0.5$.  

The final fitting curves are shown in panel (b) of Fig.~\ref{fig:fig3}, while the corresponding estimated parameters are reported in Table~\ref{Tab:Table}. For completeness, we also report the Akaike Information Criterion (AIC)~\cite{Akaike1973, hastie2017} and the Bayesian Information Criterion (BIC)~\cite{Schwarz1978, hastie2017} for the two models. We obtain $\mathrm{AIC} \approx 6011$ and $\mathrm{BIC} \approx 6042$ for the GMM, whereas $\mathrm{AIC} \approx 5127$ and $\mathrm{BIC} \approx 5159$ for the BMM. Accordingly, the Beta mixture model provides the preferred fit to the empirical data, as also visually confirmed by the corresponding fitting curve.

The fitted BMM curve  clearly highlights the bimodal nature of the data. This finding is in remarkable agreement with the arcsine distribution theoretically predicted by our probabilistic model based on the one-dimensional circular structure of the high-dimensional data. Therefore, to some extent, our directional-statistics approach supports the previous analysis based on $t$-SNE and persistence diagrams.

\begin{table}[ht]
\centering
\caption{Estimated parameters of the two-component BMM and GMM fitted to the empirical pairwise cosine-distance data. The parameters $a$ and $b$ correspond to the Beta probability density function, whereas $\mu$ and $\sigma$ denote the mean and standard deviation of the Normal distribution, respectively. }
\begin{tabular}{lccccc}
\hline
\textbf{Model} & \textbf{Comp.} & \textbf{Weight} & \textbf{Param 1} & \textbf{Param 2} \\
\hline
Beta Mixture 
 & 1 & 0.53 & $a=1.27$ & $b=3.55$ \\
Beta Mixture 
 & 2 & 0.46 & $a=4.28$ & $b=1.28$ \\
\hline
Gaussian Mixture 
 & 1 & 0.56 & $\mu=0.52$ & $\sigma=0.33$ \\
Gaussian Mixture 
 & 2 & 0.43 & $\mu=1.60$ & $\sigma=0.25$ \\
\hline
\label{Tab:Table}
\end{tabular}
\vspace{0.3cm}
\end{table}

\section{Conclusion}

Overall, we presented a possible methodology for assessing the reliability of two-dimensional embeddings produced by principal component analysis, regardless of the amount of explained variance. Our results show that when data lie on a low-dimensional nonlinear manifold, PCA may fail to provide structure-preserving scatter plots, even when the first two principal components explain a large fraction (e.g., $95\%$) of the total variance. The proposed approach is particularly suitable for relatively low-dimensional data (e.g., $n \lesssim 10$) with small sample size~\cite{damrich2024}.

Nevertheless, despite its limitations, PCA remains a valuable tool. First, it plays a crucial role in the informative initialization of $t$-SNE. Second, it can be exploited to efficiently obtain an upper bound on the intrinsic dimensionality of the data.

Furthermore, this work highlights the importance of the choice of metric in revealing the underlying data structure~\cite{chazal2021}. In our case, the Euclidean distance is sensitive to measurement noise and also fails to capture the circular symmetry of the data. In contrast, the cosine distance is well-suited for this purpose. Using the cosine distance, both $t$-SNE and persistent homology operate more reliably, yielding consistent results.

Finally, regarding the interpretation of the PCA scatter plot, Ref.~\cite{Joliffe2016} attributes the observed pattern to biological variation, suggesting that the relatively compact cluster in the lower half represents a species of Kuehneotherium, whereas the broader upper group corresponds to a related, as yet unidentified animal. In this regard, we refrain from assigning biological significance to the ring-like structure observed here, since a small set of linear measurements ($n = 9$) is unlikely to capture biologically meaningful variation in tooth shape. Modern approaches instead model the entire shape as a continuous structure, enabling the detection of subtle, taxonomically informative differences while avoiding the information loss and subjectivity associated with simpler proxies such as landmarks~\cite{ZANOLLI2023}. Accordingly, the small dataset of Kuehneotherium molar fossil teeth considered here exhibits an interesting low-dimensional ring-like structure, supported by a combined analysis based on $t$-SNE, persistent homology, and, to some extent, our directional-statistics approach, which predicts an arcsine-like distribution for the pairwise cosine distances between samples. However, this structure likely arises primarily from the generic constraints imposed by linear measurements.

~\vspace{1cm}

\begin{acknowledgments}
The author thanks Pamela Gill (University of Bristol) for kindly providing the fossil teeth measurement data used in this study and for valuable correspondence regarding the specimens and their context. The fossil teeth of Kuehneotherium are from the collections of the Natural History Museum, London, which is gratefully acknowledged. The author also thanks David M. Alba (Institut Català de Paleontologia Miquel Crusafont, Universitat Autònoma de Barcelona) for providing Ref.~\cite{ZANOLLI2023} on modern approaches to the measurement of fossil tooth shape. Finally, the author thanks Umberto Lupo for suggesting the giotto-tda library~\cite{tauzin2020giottotda} and Bastian Grossenbacher-Rieck for helpful correspondence on persistent homology diagrams and bottleneck distance.
\end{acknowledgments}



\widetext
\clearpage 
~\vspace{2cm} 
\begin{center}
\textbf{\large Supplemental Material}
\end{center}

\setcounter{equation}{0}
\setcounter{figure}{0}
\setcounter{table}{0}
\setcounter{page}{1}
\makeatletter

\renewcommand{\thefigure}{S\arabic{figure}}
\renewcommand{\thetable}{S\arabic{table}}
\renewcommand{\theequation}{S\arabic{equation}}

The supplementary materials provide information on the fossil teeth data, Andrews curves,  PCA, and $t$-SNE, including additional results.

~\vspace{1cm}

\section{Dataset and Data Preprocessing}\label{dataset}

The data come from Pamela Gill's PhD thesis (2004) on Kuehneotherium, one of the earliest known mammals~\cite{Gill2004Sup, Gill2014Sup}. The specimens date to the Early Jurassic and consist predominantly of teeth and jaws. All measurements are in millimetres and were originally taken to perform a PCA using PAST software. This dataset consists of lower molar teeth. The fossils were recovered from Mesozoic fissure fillings exposed through quarrying operations, specifically at Pant Quarry in Glamorgan, South Wales. Despite their small size, the teeth are of considerable palaeontological significance. Kuehneotherium represents one of the very earliest mammals, living some $200$ million years ago. It already possessed traits shared with modern mammals, such as two successive sets of teeth and likely a fur covering, yet it also retained more primitive features in its jaw structure and ear anatomy~\cite{Gill2025Corr}.

For computational convenience, the fossil teeth data is stored in an $n_s \times n$ matrix, where $n_s$ and $n$ denote the number of samples and variables, respectively. The preprocessing consists of standardizing the variables so that they are scale-free and have zero mean. We denote by $X^{\ast}$ the resulting standardized data matrix. Consequently, the sample covariance matrix (or, equivalently, the correlation matrix) $S$ can be constructed from $X^{\ast}$.

\section{Andrews Curves}\label{Andrews}

To begin, we search for patterns in the high-dimensional fossil teeth data using Andrews plots. The corresponding curves are shown in panel a) of Figure~\ref{fig:fig1}. We observe that the Andrews curves do not form bands, as would be expected if the data were grouped into clusters. Furthermore, the lowest curves, corresponding to labels $85, 86, 87$, and $88$, are very close to one another. Therefore, regardless of the specific manifold learning algorithm used, the two-dimensional embeddings should preserve this proximity. This is indeed the case.

We briefly present the tools used for our analysis of fossil teeth data. We start with Andrews curves~\cite{Andrews1972Sup, garcia2005Sup}. They are a useful tool for visualizing high-dimensional data, possibly revealing clustering and outliers. In this context,  each data point $\mathbf{x}= \left(x_1, x_2, \cdots, x_9 \right)$ is assigned to a  function $f(t)$ belonging to the Hilbert space   $L^{2}\left(-\pi, \pi\right)$ according to~\cite{Andrews1972Sup}
\begin{equation}
\begin{aligned}
f(t) ={}& \frac{x_1}{\sqrt{2}} + x_2 \sin(t) + x_3 \cos(t) \\
        & + x_4 \sin(2t) + x_5 \cos(2t) \\
        & + \cdots \\
        & + x_8 \sin(4t) + x_9 \cos(4t) \, .
\end{aligned}
\end{equation}

Among the nice properties of this function representation, we recall that it preserves the distances, that is, given two data points $\mathbf{x}$ and $\mathbf{y}$, and $f$ and $g$, being their representation, one finds $ \pi \lVert \mathbf{x} - \mathbf{y} \rVert^{2}_2 = \lVert f\left(t\right) - g\left(t\right) \rVert_{L^{2}}$ where $  \lVert \cdot \rVert$ and  $\lVert \cdot \rVert_{L^{2}}$ denote the Euclidean and  $L^{2}$ norms, respectively. We implemented Andrews Curves through the Pandas library~\cite{mckinney2010Sup}.

\begin{figure}
\includegraphics[width=0.80\textwidth]{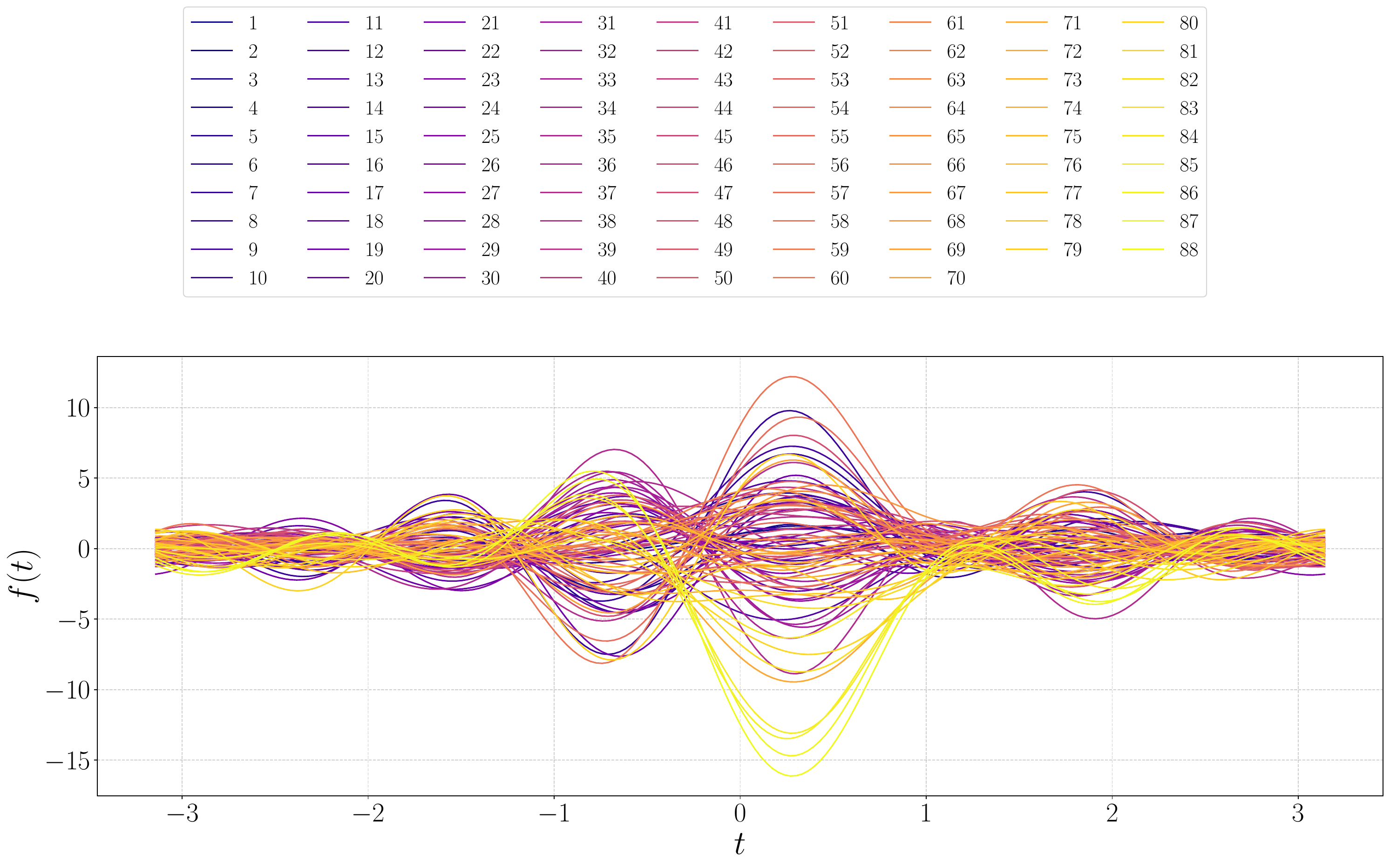}
\caption{Andrews Curves: each curve corresponds to a $9$-dimensional data point belonging to the fossil teeth dataset, consisting of $n_s= 88$ samples.}
\label{fig:fig1_SM}
\end{figure}

In Fig.~\ref{fig:fig1_SM}, the Andrews curves of the fossil tooth data, obtained using the Pandas library~\cite{mckinney2010Sup}, are shown. They provide two important insights. First, there are no evident bands, suggesting that the data should not exhibit clustering. Second, the data points with labels $85$, $86$, $87$, and $88$ appear to be close to one another. As a result, this proximity should be preserved in any corresponding low-dimensional embedding, regardless of the chosen manifold learning algorithm.

\begin{figure*}
    \centering
    \begin{minipage}{0.32\textwidth}
        \centering
        \includegraphics[width=\linewidth]{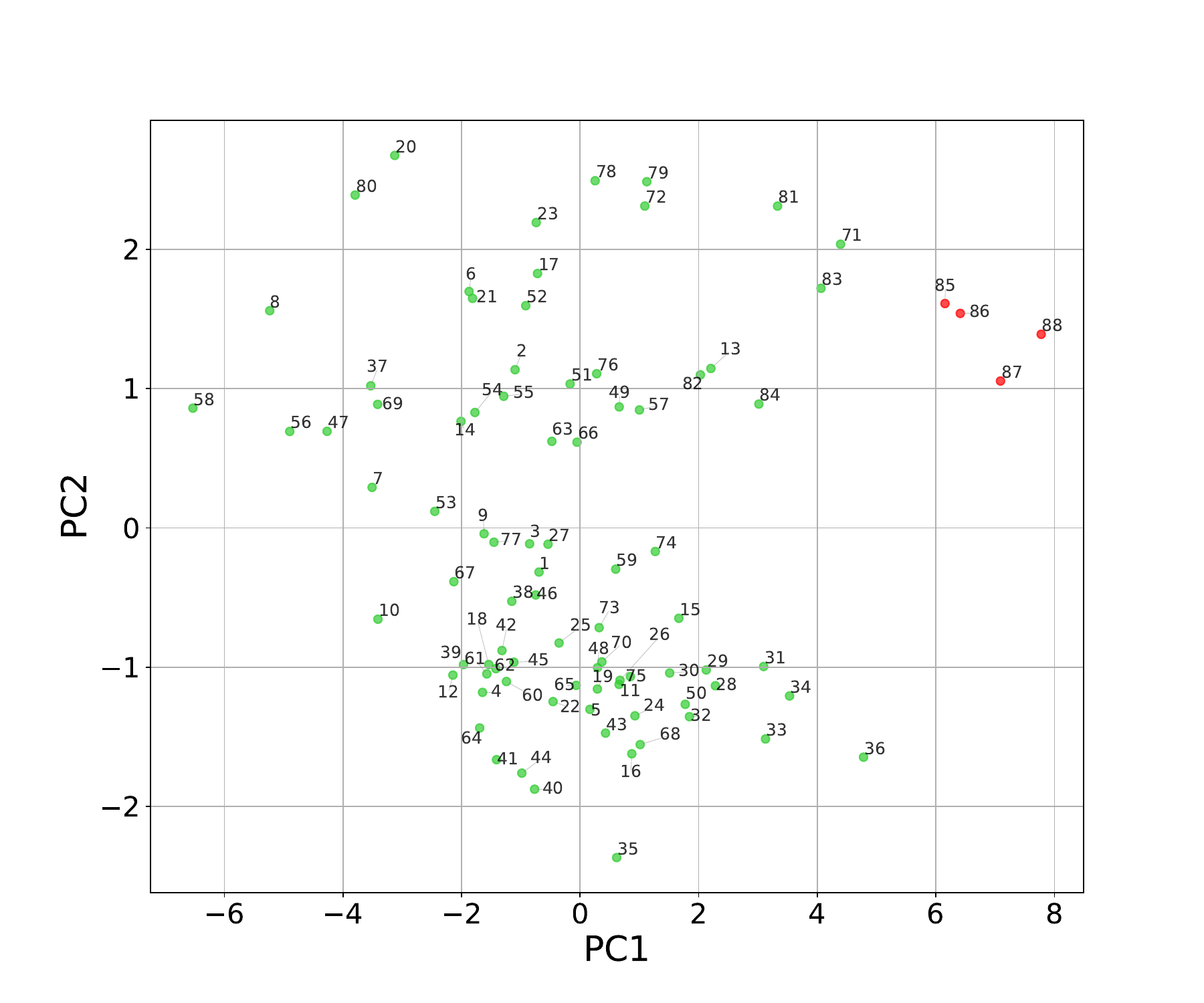}
        \caption{Scatterplot of fossil teeth data obtained by projecting the data on the optimal two-dimensional subspace. To reproduce Jolliffe and Cadima’s scatterplot faithfully, it may be necessary to swap the principal component axes, since PCA is defined only up to sign.}
        \label{fig:fig2_SM}
    \end{minipage}
    \hfill
    \begin{minipage}{0.32\textwidth}
        \centering
        \includegraphics[width=\linewidth]{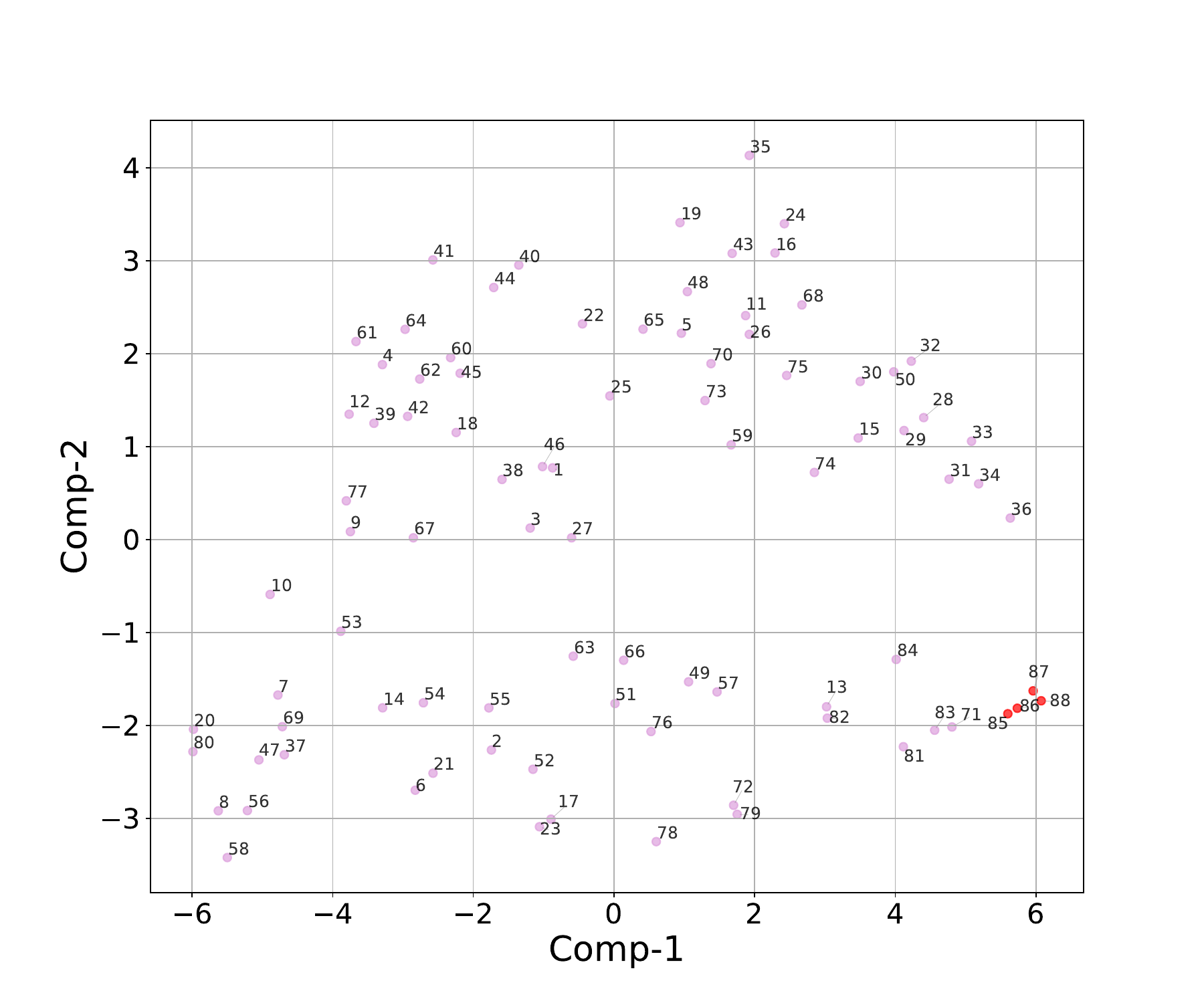}
        \caption{Scatterplot of $t$-SNE two-dimensional embedding using Euclidean distance ($p=29$). PCA was used for initialization.}
        \label{fig:fig3_SM}
    \end{minipage}
    \hfill
    \begin{minipage}{0.32\textwidth}
        \centering
        \includegraphics[width=\linewidth]{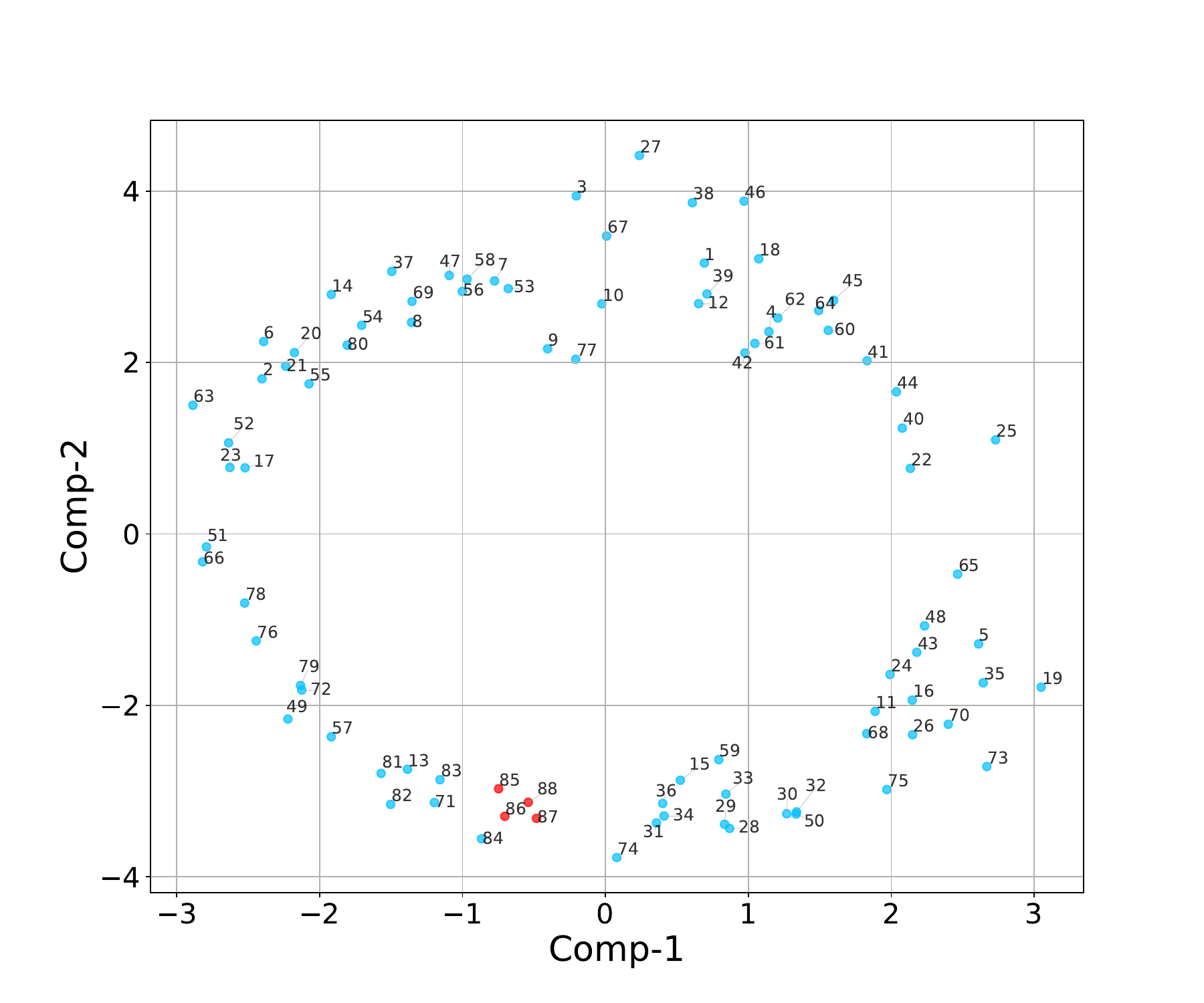}
        \caption{Scatterplot of $t$-SNE two-dimensional embedding using Euclidean distance ($p=29$). PCA was used for initialization.}
        \label{fig:fig4_SM}
    \end{minipage}
\end{figure*}

This observation is confirmed by the scatterplots corresponding to PCA and $t$-SNE (see Fig.~\ref{fig:fig2_SM}). The $t$-SNE embeddings, computed using Euclidean and cosine distances, and taking PCA as informative initialization, are shown in Fig.~\ref{fig:fig3_SM} and Fig.~\ref{fig:fig4_SM}, respectively. In all cases, the corresponding data points (highlighted in red) remain close, irrespective of the chosen method.

However, in contrast, the PCA scatterplot—which reproduces the one reported in Ref.~\cite{Jolliffe2016Sup}—exhibits a clustering tendency in the region of the linear subspace for which $\mathrm{PC2} < 0$.

Instead, both $t$-SNE embeddings are consistent with the pattern observed in the Andrews curves. However, the embedding in Fig.~\ref{fig:fig3_SM}, obtained using the Euclidean distance, does not exhibit the ring-like structure observed when using the cosine distance (see Fig.~\ref{fig:fig4_SM}). The primary reason is that noisy measurements and high dimensionality adversely affect the performance of the Euclidean metric~\cite{DamrichEtAl2023Sup, Vershynin2018Sup}.


\section{Gavish-Donoho Optimal Hard Threshold}\label{Gavish-Donoho}

\begin{figure*}
\centering
\includegraphics[width=\textwidth]{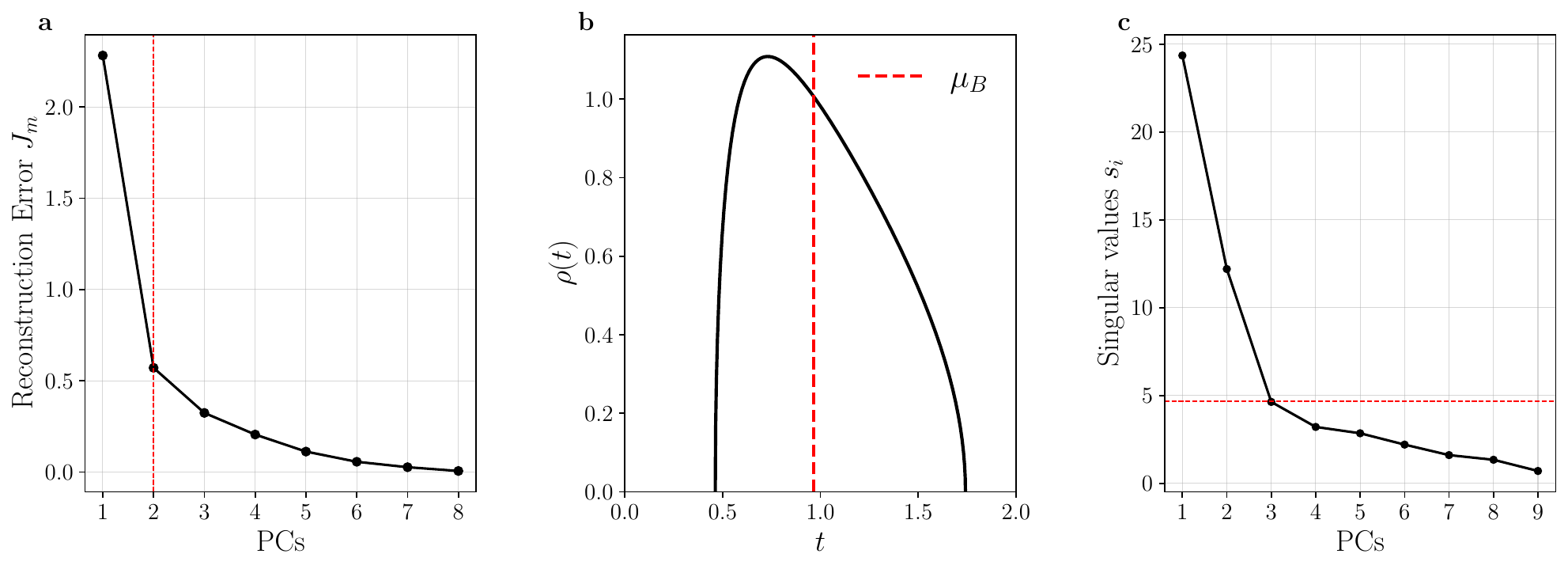}
\caption{
\textbf{a})~Reconstruction error $J_m$ as function of principal components. The vertical dashed line intersects the $J_m$ curve at the elbow point. 
\textbf{b})~ Mar\v{c}enko--Pastur  distribution $\rho\left(t\right)$. The vertical (red) dashed line intersects  MP distribution at its median $\mu_B \approx 0.96$. 
\textbf{c})~Singular values $s_i$ from SVD as function of principal components. The horizontal dashed line corresponds to the Gavish-Donoho optimal hard threshold $\tau^{\ast} \approx 4.75$ for the singular values.}
\label{fig:fig5_SM}
\end{figure*}

In the present case, considering that standardized data matrix $X^{\ast}$ has aspect ratio $\beta = n/n_s\approx 0.1$ and the noise affecting the data is unknown, $\tau^{\ast}$ can be written as~\cite{Brunton2019Sup} 

\begin{equation}
\tau^{\ast} = \omega\left(\beta\right) s_{\rm med} \, ,
\end{equation}
where $\omega \left(\beta\right) = \lambda \left(\beta\right) /\mu_B$ is the product of following function $\lambda$
\begin{equation}
\lambda\left(\beta\right) = \sqrt{
2(\beta + 1) +
\frac{8\beta}{(\beta + 1) + \sqrt{\beta^2 + 14\beta + 1}}
} \, , 
\end{equation} 
with the quantity $\mu_B$ which is the median of the Mar\v{c}enko--Pastur distribution $\rho \left(t\right)$ 
\begin{equation}
\rho(t) =
\begin{cases}
\displaystyle
\frac{1}{2\pi \beta t}
\sqrt{(\lambda_+ - t)(t - \lambda_-)},
& t \in [\lambda_-, \lambda_+] \\
\\
0, & \text{otherwise}
\end{cases}
\end{equation}
where $\lambda_{\pm} = (1 \pm \sqrt{\beta})^2$. The MP distribution for the problem at hand is shown in panel b of 
Fig.~\ref{fig:fig5_SM}.

Next, the median $\mu_B$ needs to be computed numerically by solving the following integral
\begin{equation}
\int_{\lambda_-}^{\mu_B}
\rho(t)\, d\,t = \frac{1}{2} \, .
\end{equation}

In panels a and c of Fig.~\ref{fig:fig5_SM}, the elbow method is compared to the Gavish–Donoho optimal hard threshold $\tau^{\ast} \approx 4.75$ applied to singular values $s_i$. 
Both predict $m^{\ast} = 2$ for the fossil teeth data.

\section{\texorpdfstring{$t$-SNE}{t-SNE} Details}\label{tSNE_details}

The symmetric probabilities $p_{ij} = \left(2 n_s \right)^{-1} \left(p_{i\mid j} + p_{j\mid i}\right)$  and $q_{ij} =  \left(2 n_s \right)^{-1}  \left(q_{i\mid j} + q_{j\mid i}\right)$, entering into Eq.~\ref{eq:KL}, depend on the conditional probabilities $p_{j\mid i}$ and $p_{j\mid i}$, respectively. The probabilities $p_{ij}$ and $q_{ij}$ measure the similarity between $\mathbf{x_i}$, $\mathbf{x_j}$ and $\mathbf{y_i}$, $\mathbf{y_j}$, respectively.

The probability $p_{j\mid i}$ yields the probability that $\mathbf{x_j}$ would be a neighbor of $\mathbf{x_i}$, as a Gaussian kernel:
\begin{equation}\label{eq:conditional_p}
	p_{j\mid i} = \frac{\exp(- \lVert \mathbf{x_i} - \mathbf{x_j} \rVert_2^{2} / 2\sigma_i^2)}{\sum_{k=1, k \neq i}^{n_s} \exp(-\lVert \mathbf{x_i} - \mathbf{x_k} \rVert_2^{2} / 2\sigma_i^2)}  \, ,
\end{equation}
where the width $\sigma_i$  of the kernel represents the local density. 

 The variance $\sigma_i^2$ is estimated by specifying the parameter $\tau_p$. The latter is understood as the effective number of neighbors.

Finally, $q_{j\mid i}$ gives the probability that $\mathbf{y}_j$ would be a neighbor of $\mathbf{y}_i$. However, in contrast to $p_{j\mid i}$,  the probability $q_{ij}$ rests  on the t-distribution with one degree of freedom (the Cauchy distribution) and reads
\begin{equation}\label{eq:t_student}
	q_{i j}=\frac{\left(1+\left\|\mathbf{y}_{i}-\mathbf{y}_{j}\right\|_2^{2}\right)^{-1}}{\sum_{k =1, k \neq l}^{n_s}\left(1+\left\|\mathbf{y}_{k}-\mathbf{y}_{l}\right\|_2^{2}\right)^{-1}}  \, .
\end{equation}

\section{Persistent Homology Barcodes}\label{extra_results}

A persistent homology barcode provides a distinct, yet equivalent, representation of the persistence diagram of a dataset, thereby encoding the lifetimes of topological features across different scales. In the present case, the scale parameter is given by the diameter of the Vietoris–Rips complex, which in turn depends on the chosen metric.

\begin{figure*}
    \centering
    \begin{minipage}{0.45\textwidth}
        \centering
        \includegraphics[width=\linewidth]{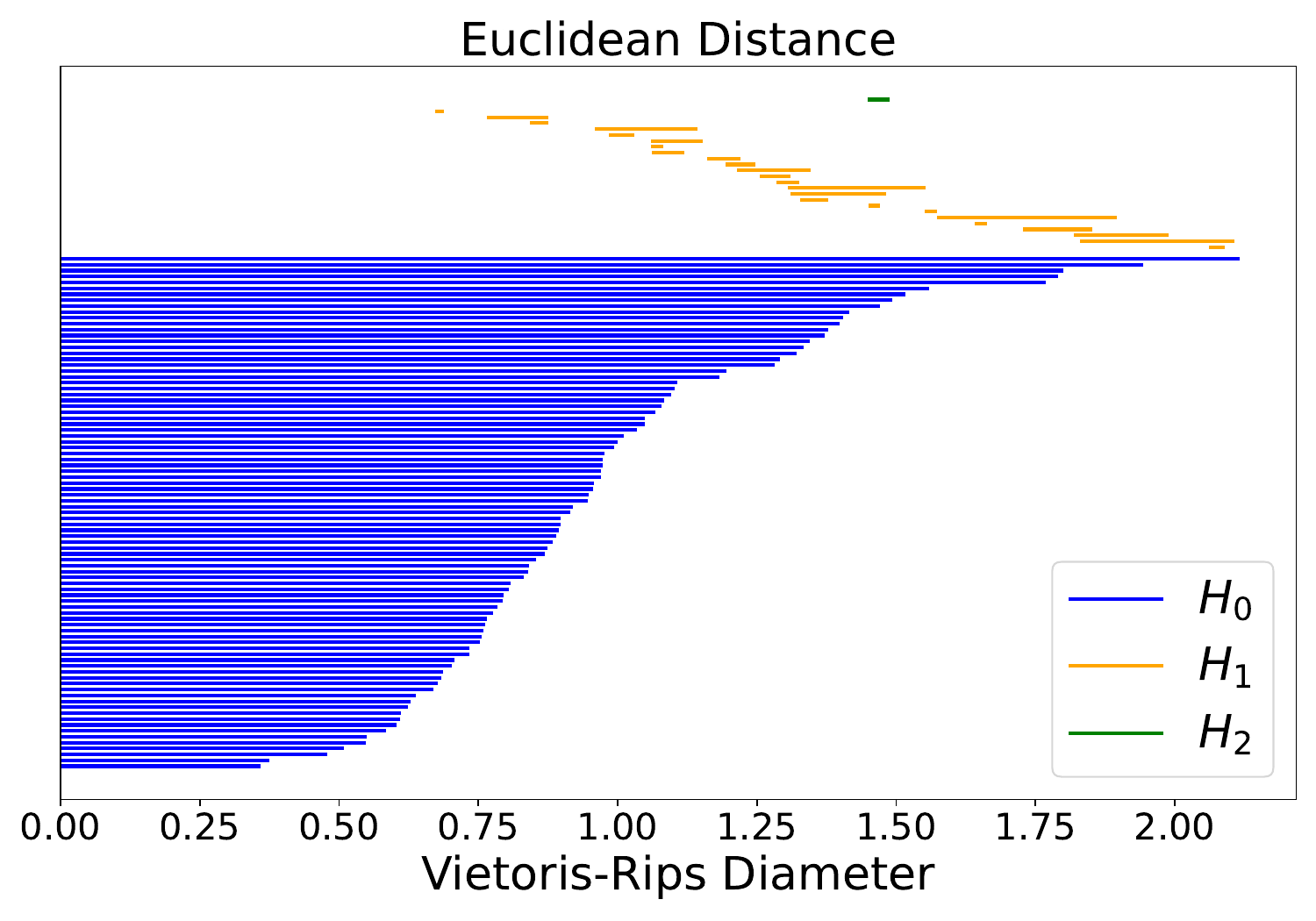}
        \caption{Barcodes using Vietoris–Rips simplicial complex and Euclidean distance. Solid lines as functions of Vietoris-Rips diameter represent the lifetimes of connected components ($H_0$), holes  ($H_1$), and
        voids ($H_2$).}
        \label{fig:fig6_SM}
    \end{minipage}
    \hfill
    \begin{minipage}{0.45\textwidth}
        \centering
        \includegraphics[width=\linewidth]{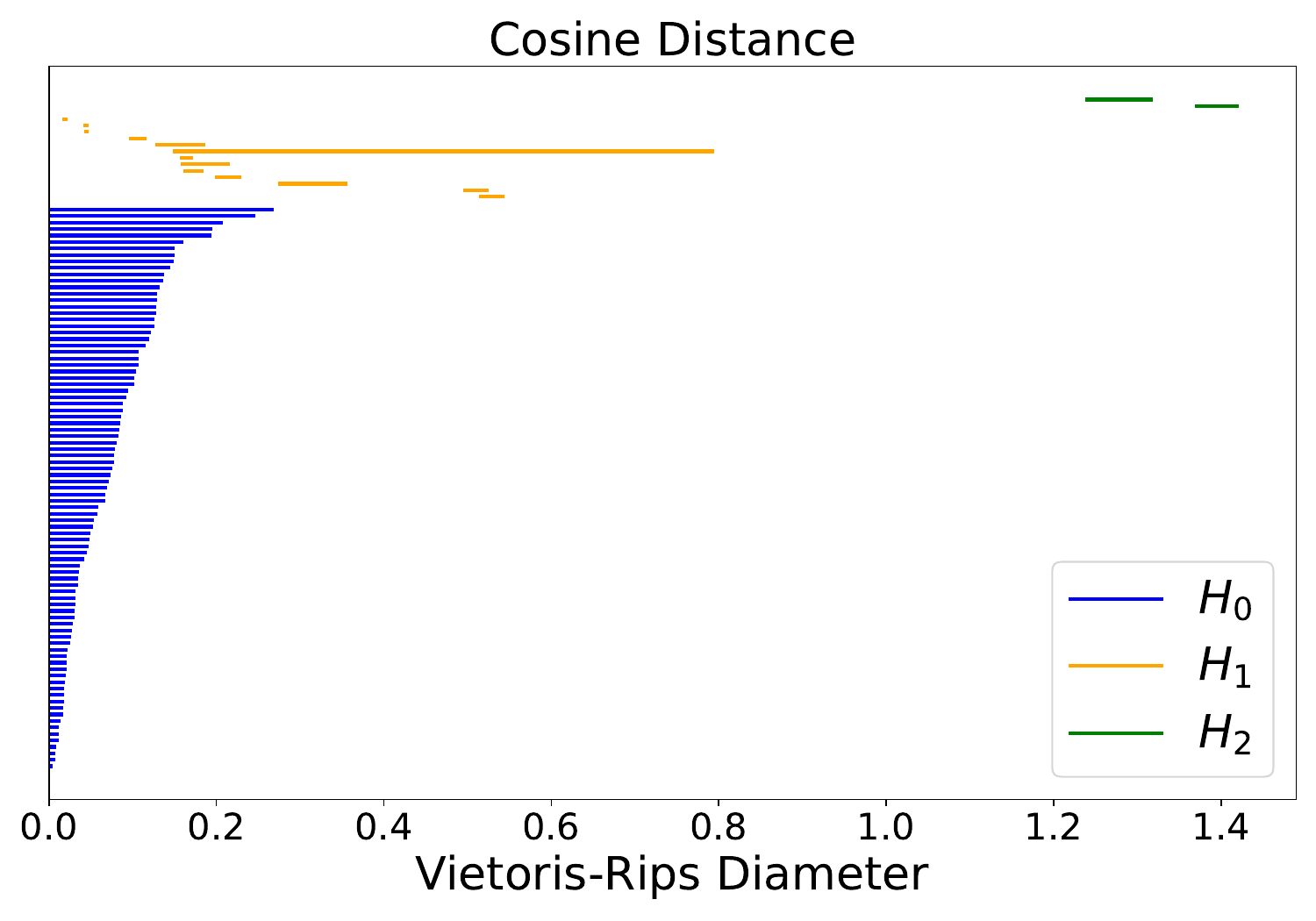}
        \caption{Barcodes using Vietoris–Rips simplicial complex and cosine distance. Solid lines as functions of Vietoris-Rips diameter represent the lifetimes of connected components ($H_0$), holes  ($H_1$), and
        voids ($H_2$).}
        \label{fig:fig7_SM}
    \end{minipage}
\end{figure*}

Figures~\ref{fig:fig6_SM} and~\ref{fig:fig7_SM} show the barcodes of the fossil teeth data computed using the Euclidean and cosine distances, respectively.

Regarding $H_0$, both barcodes display $86$ bars, all originating at $0$, corresponding to initially isolated points. Under the Euclidean distance, the connected components merge over a wider range of scales (death values up to $\approx 2.1$) than in the cosine distance case (death values up to $\approx 0.27$). This difference reflects the distinct scales induced by the two metrics. In the cosine distance barcode, the earlier merging of $H_0$ components results from points having nearby angular neighbors, consistent with a dense distribution along the ring-like structure. Nevertheless, both $H_0$ patterns indicate that the data form a fragmented cloud that merges progressively, rather than well-separated clusters.

Turning to the $H_1$ bars obtained using the Euclidean distance, most are short-lived, suggesting transient, noise-driven circular structures. However, a few bars exhibit longer lifetimes. In particular, the bar born at $\approx 1.57$ and dying at $\approx 1.89$ has persistence $\approx 0.32$, thereby providing partial evidence for the presence of a loop in the data. In contrast, the barcode computed using the cosine distance contains a markedly longer-lived $H_1$ bar, with persistence $\approx 0.64$, born at $\approx 0.15$ and dying at $\approx 0.79$. This strongly suggests the presence of a genuine ring-like structure in the data. Accordingly, the corresponding hole cannot be filled without a substantial increase in the Vietoris–Rips diameter, giving rise to the observed long-lived $H_1$ feature.

Finally, the short-lived $H_2$ bars observed in both barcodes are consistent with topological noise.

\bibliography{aipsamp}

@PREAMBLE{
 "\providecommand{\noopsort}[1]{}" 
 # "\providecommand{\singleletter}[1]{#1}%" 
}

@Article{Gill2014,
author={Gill, Pamela G.
and Purnell, Mark A.
and Crumpton, Nick
and Brown, Kate Robson
and Gostling, Neil J.
and Stampanoni, M.
and Rayfield, Emily J.},
title={Dietary specializations and diversity in feeding ecology of the earliest stem mammals},
journal={Nature},
year={2014},
month={Aug},
day={01},
volume={512},
number={7514},
pages={303-305},
issn={1476-4687},
doi={10.1038/nature13622},
url={https://doi.org/10.1038/nature13622}
}

@article{Shinn2023,
  author  = {Shinn, Maxwell},
  title   = {Phantom Oscillations in Principal Component Analysis},
  journal = {Proceedings of the National Academy of Sciences},
  year    = {2023},
  volume  = {120},
  number  = {48},
  pages   = {e2311420120},
  doi     = {10.1073/pnas.2311420120}
}

@article{ZANOLLI2023,
title = {A reassessment of the distinctiveness of dryopithecine genera from the Iberian Miocene based on enamel-dentine junction geometric morphometric analyses},
journal = {Journal of Human Evolution},
volume = {177},
pages = {103326},
year = {2023},
issn = {0047-2484},
doi = {https://doi.org/10.1016/j.jhevol.2023.103326},
url = {https://www.sciencedirect.com/science/article/pii/S0047248423000039},
author = {Clément Zanolli and Florian Bouchet and Josep Fortuny and Federico Bernardini and Claudio Tuniz and David M. Alba}
}

@misc{debodt2025,
      title={Low-dimensional embeddings of high-dimensional data}, 
      author={Cyril de Bodt and Alex Diaz-Papkovich and Michael Bleher and Kerstin Bunte and Corinna Coupette and Sebastian Damrich and Enrique Fita Sanmartin and Fred A. Hamprecht and Emőke-Ágnes Horvát and Dhruv Kohli and Smita Krishnaswamy and John A. Lee and Boudewijn P. F. Lelieveldt and Leland McInnes and Ian T. Nabney and Maximilian Noichl and Pavlin G. Poličar and Bastian Rieck and Guy Wolf and Gal Mishne and Dmitry Kobak},
      year={2025},
      eprint={2508.15929},
      archivePrefix={arXiv},
      primaryClass={cs.LG},
      url={https://arxiv.org/abs/2508.15929}, 
}

@book{Bury1999,
  author    = {Bury, Karl V.},
  title     = {Statistical Distributions in Engineering},
  publisher = {Cambridge University Press},
  year      = {1999}
}

@book{MardiaJupp2000,
  author    = {Mardia, Kanti V. and Jupp, Peter E.},
  title     = {Directional Statistics},
  publisher = {Wiley},
  year      = {2000},
  series    = {Wiley Series in Probability and Statistics},
  isbn      = {978-0471953333}
}

@book{Casella2002,
  author    = {Casella, George and Berger, Roger L.},
  title     = {Statistical Inference},
  edition   = {2},
  publisher = {Duxbury},
  year      = {2002}
}

@article{Levy1939,
  author  = {L{\'e}vy, Paul},
  title   = {Sur certains processus stochastiques homog{\`e}nes},
  journal = {Compositio Mathematica},
  volume  = {7},
  pages   = {283--339},
  year    = {1939}
}

@book{Vershynin2018,
  author    = {Vershynin, Roman},
  title     = {High-Dimensional Probability: An Introduction with Applications in Data Science},
  publisher = {Cambridge University Press},
  year      = {2018},
  series    = {Cambridge Series in Statistical and Probabilistic Mathematics},
  doi       = {10.1017/9781108231596}
}

@Article{Horton2025,
author={Horton, Matthew K.
and Huck, Patrick
and Yang, Ruo Xi
and Munro, Jason M.
and Dwaraknath, Shyam
and Ganose, Alex M.
and Kingsbury, Ryan S.
and Wen, Mingjian
and Shen, Jimmy X.
and Mathis, Tyler S.
and Kaplan, Aaron D.
and Berket, Karlo
and Riebesell, Janosh
and George, Janine
and Rosen, Andrew S.
and Spotte-Smith, Evan W. C.
and McDermott, Matthew J.
and Cohen, Orion A.
and Dunn, Alex
and Kuner, Matthew C.
and Rignanese, Gian-Marco
and Petretto, Guido
and Waroquiers, David
and Griffin, Sinead M.
and Neaton, Jeffrey B.
and Chrzan, Daryl C.
and Asta, Mark
and Hautier, Geoffroy
and Cholia, Shreyas
and Ceder, Gerbrand
and Ong, Shyue Ping
and Jain, Anubhav
and Persson, Kristin A.},
title={Accelerated data-driven materials science with the Materials Project},
journal={Nature Materials},
year={2025},
month={Oct},
day={01},
volume={24},
number={10},
pages={1522-1532},
issn={1476-4660},
doi={10.1038/s41563-025-02272-0},
url={https://doi.org/10.1038/s41563-025-02272-0}
}

@article{Zeng_2024,
doi = {10.1088/2632-2153/ad4ba5},
url = {https://doi.org/10.1088/2632-2153/ad4ba5},
year = {2024},
month = {may},
publisher = {IOP Publishing},
volume = {5},
number = {2},
pages = {025053},
author = {Zeng, Kevin and De Jesús, Carlos E Pérez and Fox, Andrew J and Graham, Michael D},
title = {Autoencoders for discovering manifold dimension and coordinates in data from complex dynamical systems},
journal = {Machine Learning: Science and Technology}

}

@book{Villani2008,
  author    = {Villani, C{\'e}dric},
  title     = {Optimal Transport: Old and New},
  publisher = {Springer},
  address   = {Berlin, Heidelberg},
  year      = {2008}
}

@article{PeyreCuturi2018,
  author  = {Peyr{\'e}, Gabriel and Cuturi, Marco},
  title   = {Computational Optimal Transport},
  journal = {arXiv preprint arXiv:1803.00567},
  year    = {2018}
}

@article{Wasserman2018,
  author  = {Wasserman, Larry},
  title   = {Topological Data Analysis},
  journal = {Annual Review of Statistics and Its Application},
  year    = {2018},
  volume  = {5},
  number  = {1},
  pages   = {501--532},
  doi     = {10.1146/annurev-statistics-031017-100045}
}

@misc{tauzin2020giottotda,
      title={giotto-tda: A Topological Data Analysis Toolkit for Machine Learning and Data Exploration},
      author={Guillaume Tauzin and Umberto Lupo and Lewis Tunstall and Julian Burella Pérez and Matteo Caorsi and Anibal Medina-Mardones and Alberto Dassatti and Kathryn Hess},
      year={2020},
      eprint={2004.02551},
      archivePrefix={arXiv},
      primaryClass={cs.LG}
}

@misc{damrich2024,
      title={Persistent Homology for High-dimensional Data Based on Spectral Methods}, 
      author={Sebastian Damrich and Philipp Berens and Dmitry Kobak},
      year={2024},
      eprint={2311.03087},
      archivePrefix={arXiv},
      primaryClass={cs.LG},
      url={https://arxiv.org/abs/2311.03087}, 
}

@misc{DamrichEtAl2023Sup,
      title={Persistent Homology for High-dimensional Data Based on Spectral Methods}, 
      author={Sebastian Damrich and Philipp Berens and Dmitry Kobak},
      year={2024},
      eprint={2311.03087},
      archivePrefix={arXiv},
      primaryClass={cs.LG},
      url={https://arxiv.org/abs/2311.03087}, 
}

@article{strang1993,
 author = {Gilbert Strang},
 journal = {The American Mathematical Monthly},
 number = {9},
 pages = {848--855},
 publisher = {[Taylor & Francis, Ltd., Mathematical Association of America]},
 title = {The Fundamental Theorem of Linear Algebra},
 urldate = {2024-08-06},
 volume = {100},
 year = {1993}
}

@article{stewart1993,
author = {Stewart, G. W.},
title = {On the Early History of the Singular Value Decomposition},
journal = {SIAM Review},
volume = {35},
number = {4},
pages = {551-566},
year = {1993}
}

@article{pedregosa2011,
  author  = {Fabian Pedregosa and Ga{{\"e}}l Varoquaux and Alexandre Gramfort and Vincent Michel and Bertrand Thirion and Olivier Grisel and Mathieu Blondel and Peter Prettenhofer and Ron Weiss and Vincent Dubourg and Jake Vanderplas and Alexandre Passos and David Cournapeau and Matthieu Brucher and Matthieu Perrot and {{\'E}}douard Duchesnay},
  title   = {Scikit-learn: Machine Learning in Python},
  journal = {Journal of Machine Learning Research},
  year    = {2011},
  volume  = {12},
  number  = {85},
  pages   = {2825--2830},
  url     = {http://jmlr.org/papers/v12/pedregosa11a.html}
}

@InProceedings{mckinney-proc-scipy-2010,
  author    = { {W}es {M}c{K}inney },
  title     = { {D}ata {S}tructures for {S}tatistical {C}omputing in {P}ython },
  booktitle = { {P}roceedings of the 9th {P}ython in {S}cience {C}onference },
  pages     = { 56 - 61 },
  year      = { 2010 },
  editor    = { {S}t\'efan van der {W}alt and {J}arrod {M}illman },
  doi       = { 10.25080/Majora-92bf1922-00a }
}

@book{Brunton2019,
  author    = {Brunton, Steven L. and Kutz, J. Nathan},
  title     = {Data-Driven Science and Engineering: Machine Learning, Dynamical Systems, and Control},
  publisher = {Cambridge University Press},
  year      = {2019},
  doi       = {10.1017/9781108380690}
}

@article{policar2024,
 title={openTSNE: A Modular Python Library for t-SNE Dimensionality Reduction and Embedding},
 volume={109},
 url={https://www.jstatsoft.org/index.php/jss/article/view/v109i03},
 doi={10.18637/jss.v109.i03},
 journal={Journal of Statistical Software},
 author={Poličar, Pavlin G. and Stražar, Martin and Zupan, Blaž},
 year={2024},
 pages={1–30}
}

@article{Akaike1973,
  author  = {Akaike, Hirotugu},
  title   = {Information Theory and an Extension of the Maximum Likelihood Principle},
  journal = {2nd International Symposium on Information Theory},
  year    = {1973},
  pages   = {267--281}
}

@article{Schwarz1978,
  author    = {Schwarz, Gideon},
  title     = {Estimating the Dimension of a Model},
  journal   = {The Annals of Statistics},
  year      = {1978},
  volume    = {6},
  number    = {2},
  pages     = {461--464},
  doi       = {10.1214/aos/1176344136}
}

@book{Murphy2012,
	address = {Cambridge, Massachusetts},
	author = {Murphy, K.~ P.},
	publisher = {MIT Press},
	title = {Machine learning - a probabilistic perspective},
	year = 2012}

@article{scheidgen2023nomad,
  author    = {Matthias Scheidgen and Lauri Himanen and Antonio Ladines and
               D{\'a}vid Sikter and Mahdieh Nakhaee and {\'A}d{\'a}m Fekete and
               Tien-Chien Chang and Amir Golparvar and Juan Mar{\'\i}quez and
               Stephan Brockhauser and Stefan Br{\"u}ckner and Luca M. Ghiringhelli and
               Fabian Dietrich and Dominik Lehmberg and Thomas Denell and
               Andrea Albino and H{\"a}kan N{\"a}str{\"o}m and Siamak Shabih and
               Felix Dobener and Michael K{\"u}hbach and Ria Mozumder and
               Jonathan Rudzinski and Nick Daelman and Jesus Pizarro and
               Markus Kuban and Christian Salazar and Peter Ondra{\v c}ka and
               Hans-Joachim Bungartz and Claudia Draxl},
  title     = {NOMAD: A distributed web-based platform for managing materials science research data},
  journal   = {Journal of Open Source Software},
  volume    = {8},
  number    = {90},
  pages     = {5388},
  year      = {2023},
  doi       = {10.21105/joss.05388},
  url       = {https://doi.org/10.21105/joss.05388}
}

@article{berman2000protein,
  author    = {Helen M. Berman and Tom Battistuz and Talapady N. Bhat and
               William Bluhm and Peter E. Bourne and Karen Burkhardt and
               Leif Iype and Sanjay Jain and Philip Fagan and J. Marvin and
               David Padilla and Veerasamy Ravichandran and Barry I. Schneider and
               Nitin Thanki and H. Weissig and John Westbrook and C. Zardecki},
  title     = {The Protein Data Bank},
  journal   = {Nucleic Acids Research},
  volume    = {28},
  number    = {1},
  pages     = {235--242},
  year      = {2000},
  publisher = {Oxford University Press},
  note      = {The worldwide repository of experimentally determined macromolecular structures},
  url       = {https://www.rcsb.org/}
}

@phdthesis{Gill2004Sup,
  author = {Gill, P. G.},
  title = {Kuehneotherium from the Mesozoic Fissure Fillings of South Wales},
  school = {University of Bristol},
  year = {2004}
}

@misc{Gill2025Corr,
  author = {Gill, P. G.},
  title = {Personal correspondence},
  year = {2025}
}

@article{Andrews1972Sup,
  author = {Andrews, D. F.},
  title = {Plots of High-Dimensional Data},
  journal = {Biometrics},
  volume = {28},
  number = {1},
  pages = {125--136},
  year = {1972},
  doi = {10.2307/2528964}
}

@article{garcia2005Sup,
  author = {Garc{\'i}a-Osorio, C. and Fyfe, C.},
  title = {Visualization of High-Dimensional Data via Orthogonal Curves},
  journal = {Journal of Universal Computer Science},
  volume = {11},
  pages = {1806--1819},
  year = {2005}
}

@article{Gill2014Sup,
  author = {Gill, P. G. and Purnell, M. A. and Crumpton, N. and Brown, K. R. and Gostling, N. J. and Stampanoni, M. and Rayfield, E. J.},
  title = {Dietary specializations and diversity in feeding ecology of the earliest stem mammals},
  journal = {Nature},
  volume = {512},
  number = {7514},
  pages = {303--305},
  year = {2014}
}

@inproceedings{mckinney2010Sup,
  author = {McKinney, Wes},
  title = {Data Structures for Statistical Computing in Python},
  booktitle = {Proceedings of the 9th Python in Science Conference},
  pages = {56--61},
  year = {2010},
  editor = {van der Walt, St{\'e}fan and Millman, Jarrod},
  doi = {10.25080/Majora-92bf1922-00a}
}

@article{Jolliffe2016Sup,
  author = {Jolliffe, Ian T. and Cadima, Jorge},
  title = {Principal component analysis: a review and recent developments},
  journal = {Philosophical Transactions of the Royal Society A: Mathematical, Physical and Engineering Sciences},
  volume = {374},
  number = {2065},
  pages = {20150202},
  year = {2016},
  doi = {10.1098/rsta.2015.0202}
}

@book{Vershynin2018Sup,
  author = {Vershynin, Roman},
  title = {High-Dimensional Probability: An Introduction with Applications in Data Science},
  publisher = {Cambridge University Press},
  year = {2018},
  series = {Cambridge Series in Statistical and Probabilistic Mathematics},
  doi = {10.1017/9781108231596}
}

@book{Brunton2019Sup,
  author = {Brunton, Steven L. and Kutz, J. Nathan},
  title = {Data-Driven Science and Engineering: Machine Learning, Dynamical Systems, and Control},
  publisher = {Cambridge University Press},
  year = {2019},
  doi = {10.1017/9781108380690}
}

@article{Kramer1993,
  author = {Kramer, Mark A.},
  title = {Nonlinear Principal Component Analysis Using Autoassociative Neural Networks},
  journal = {AIChE Journal},
  volume = {37},
  number = {2},
  pages = {233--243},
  year = {1991},
  doi = {10.1002/aic.690370209}
}

@article{varadi2024alphafold,
  author    = {Mihaly Varadi and Damian Bertoni and Paulyna Magana and Urmila Paramval and Ivanna Pidruchna and
               Malarvizhi Radhakrishnan and Maxim Tsenkov and Sreenath Nair and Milot Mirdita and Jingi Yeo and
               Oleg Kovalevskiy and Kathryn Tunyasuvunakool and Agata Laydon and Augustin {\v{Z}}{\'\i}dek and
               Hamish Tomlinson and Dhavanthi Hariharan and Josh Abrahamson and Tim Green and John Jumper and
               Ewan Birney and Martin Steinegger and Demis Hassabis and Sameer Velankar},
  title     = {AlphaFold Protein Structure Database in 2024: Providing structure coverage for over 214 million protein sequences},
  journal   = {Nucleic Acids Research},
  volume    = {52},
  number    = {D1},
  pages     = {D368--D375},
  year      = {2024},
  doi       = {10.1093/nar/gkad1011}
}

@article{Greenacre2022,
  title={Principal component analysis},
  author={Michael Greenacre and Patrick J. F. Groenen and Trevor Hastie and Alfonso Iodice D’Enza and Angelos I. Markos and Elena Tuzhilina},
  journal={Nature Reviews Methods Primers},
  year={2022},
  volume={2}
}

@article{Hotelling1933,
  title={Analysis of a complex of statistical variables into principal components.},
  author={Harold Hotelling},
  journal={Journal of Educational Psychology},
  year={1933},
  volume={24},
  pages={498-520}
}

@misc{shlens2014,
      title={A Tutorial on Principal Component Analysis}, 
      author={Jonathon Shlens},
      year={2014},
      eprint={1404.1100},
      archivePrefix={arXiv},
      primaryClass={cs.LG},
      url={https://arxiv.org/abs/1404.1100}, 
}

@article{RECANATESI2022,
title = {A scale-dependent measure of system dimensionality},
journal = {Patterns},
volume = {3},
number = {8},
pages = {100555},
year = {2022},
issn = {2666-3899},
doi = {https://doi.org/10.1016/j.patter.2022.100555},
url = {https://www.sciencedirect.com/science/article/pii/S266638992200160X},
author = {Stefano Recanatesi and Serena Bradde and Vijay Balasubramanian and Nicholas A. Steinmetz and Eric Shea-Brown}
}

@book{jolliffe2002pca,
  author       = {I. T. Jolliffe},
  title        = {Principal Component Analysis},
  series       = {Springer Series in Statistics},
  edition      = {2},
  publisher    = {Springer},
  address      = {New York, NY},
  year         = {2002},
  isbn         = {978-0-387-95442-4},
  doi          = {10.1007/b98835},
  pages        = {XXX, 488},
  note         = {Springer Science+Business Media New York; eBook ISBN: 978-0-387-22440-4; Softcover ISBN: 978-1-4419-2999-0; Published in Springer Book Archive},
}

@article{Kaiser1960,
author = {Henry F. Kaiser},
title ={The Application of Electronic Computers to Factor Analysis},

journal = {Educational and Psychological Measurement},
volume = {20},
number = {1},
pages = {141-151},
year = {1960},
doi = {10.1177/001316446002000116},
URL = {https://doi.org/10.1177/001316446002000116},
eprint = {https://doi.org/10.1177/001316446002000116}

}

@ARTICLE{Gavish2014,
  author={Gavish, Matan and Donoho, David L.},
  journal={IEEE Transactions on Information Theory}, 
  title={The Optimal Hard Threshold for Singular Values is  $4/\sqrt {3}$ }, 
  year={2014},
  volume={60},
  number={8},
  pages={5040-5053},
  doi={10.1109/TIT.2014.2323359}}

@article{marcenko1967,
  author    = {V. A. Mar\v{c}enko and L. A. Pastur},
  title     = {Distribution of eigenvalues for some sets of random matrices},
  journal   = {Mathematics of the USSR-Sbornik},
  volume    = {1},
  number    = {4},
  pages     = {457--483},
  year      = {1967}
}

@book{Geron2017,
  title     = "Hands-On Machine Learning with Scikit-Learn and TensorFlow: Concepts, Tools, and Techniques to Build Intelligent Systems",
  author    = "G{\'e}ron, Aur{\'e}lien",
  year      = 2019,
  publisher = "O'ReillY",
address   = " U.S.A"
}

@article{tenenbaum2000,
author = {Joshua B. Tenenbaum  and Vin de Silva  and John C. Langford },
title = {A Global Geometric Framework for Nonlinear Dimensionality Reduction},
journal = {Science},
volume = {290},
number = {5500},
pages = {2319-2323},
year = {2000}
}

@article{Pearson1901,
  title={On lines and planes of closest fit to systems of points in space},
  author={Karl Pearson},
  journal={Philosophical Magazine Series 1},
  year={1901},
  volume={2},
  pages={559-572}
}

@article{Munch_2017, title={A User’s Guide to Topological Data Analysis}, volume={4}, url={https://learning-analytics.info/index.php/JLA/article/view/5196}, DOI={10.18608/jla.2017.42.6}, abstractNote={&lt;p&gt;Topological data analysis (TDA) is a collection of powerful tools that can quantify shape and structure in data in order to answer questions from the data’s domain. This is done by representing some aspect of the structure of the data in a simplified topological signature. In this article, we introduce two of the most commonly used topological signatures. First, the persistence diagram represents loops and holes in the space by considering connectivity of the data points for a continuum of values rather than a single fixed value. The second topological signature, the mapper graph, returns a 1-dimensional structure representing the shape of the data, and is particularly good for exploration and visualization of the data. While these techniques are based on very sophisticated mathematics, the current ubiquity of available software means that these tools are more accessible than ever to be applied to data by researchers in education and learning, as well as all domain scientists.&lt;/p&gt;}, number={2}, journal={Journal of Learning Analytics}, author={Munch, Elizabeth}, year={2017}, month={Jul.}, pages={47–61} }

@article{vandermaaten2008,
  author  = {Laurens van der Maaten and Geoffrey Hinton},
  title   = {Visualizing Data using t-SNE},
  journal = {Journal of Machine Learning Research},
  year    = {2008},
  volume  = {9},
  number  = {86},
  pages   = {2579--2605}
}

@Article{Kobak2019,
author={Kobak, Dmitry
and Berens, Philipp},
title={The art of using t-SNE for single-cell transcriptomics},
journal={Nature Communications},
year={2019},
month={Nov},
day={28},
volume={10},
number={1},
pages={5416},
issn={2041-1723},
doi={10.1038/s41467-019-13056-x},
url={https://doi.org/10.1038/s41467-019-13056-x}
}

@article{Carlsson2009,
  title={Topology and data},
  author={Gunnar E. Carlsson},
  journal={Bulletin of the American Mathematical Society},
  year={2009},
  volume={46},
  pages={255-308}
}

@Article{Otter2017,
author={Otter, Nina
and Porter, Mason A.
and Tillmann, Ulrike
and Grindrod, Peter
and Harrington, Heather A.},
title={A roadmap for the computation of persistent homology},
journal={EPJ Data Science},
year={2017},
month={Aug},
day={09},
volume={6},
number={1},
pages={17},
issn={2193-1127},
doi={10.1140/epjds/s13688-017-0109-5},
url={https://doi.org/10.1140/epjds/s13688-017-0109-5}
}

@book{hastie2017,
 
  address = {New York, NY, USA},
  author = {Hastie, Trevor and Tibshirani, Robert and Friedman, Jerome},
 edition = {12},
  publisher = {Springer New York Inc.},
  series = {Springer Series in Statistics},
  title = {The Elements of Statistical Learning},
  year = {2017}
}

@article{Joliffe2016,
author = {Jolliffe, Ian T.  and Cadima, Jorge },
title = {Principal component analysis: a review and recent developments},
journal = {{Philosophical Transactions of the Royal Society A:
Mathematical, Physical and Engineering Sciences}},
volume = {374},
number = {2065},
pages = {20150202},
year = {2016},
doi = {10.1098/rsta.2015.0202}
}

@book{Deisenroth2020,
 
  author = {Deisenroth, Marc Peter and Faisal, A. Aldo and Ong, Cheng Soon},
  title={{ Mathematics for Machine Learning }},
  publisher = {Cambridge University Press},
  year={2020}
}

@ARTICLE{chazal2021,
    author={Chazal, Frédéric and Michel, Bertrand},    
    title={An Introduction to Topological Data Analysis: Fundamental and Practical Aspects for Data Scientists},      
    journal={Frontiers in Artificial Intelligence},      
    volume={4},           
    year={2021},
}

\end{document}